\documentclass[twocolumn,showpacs,prl]{revtex4}
\usepackage{graphicx}

\begin{document}

\title{Reply to the reply to ``Comment on Universal out-of-equilibrium transport in Kondo-correlated quantum dots''}
\author{A. A. Aligia}
\affiliation{Centro At\'{o}mico Bariloche and Instituto Balseiro, Comisi\'{o}n Nacional
de Energ\'{\i}a At\'{o}mica, 8400 Bariloche, Argentina}
\pacs{72.10.Bg, 72.15.Qm, 73.21.La, 75.30.Mb}
\date{\today}
\maketitle

Two recent works attempt to extend results for the conductance through a
quantum dot described by the particle-hole symmetric (PHS) impurity Anderson
model out of the PHS case using renormalized perturbation theory in $U$
(PTU) \cite{sca,mu}. A controversy between the authors of these papers
exists \cite{com,reply}. First Mu\~{n}oz {\it et al.} 
added an incorrect note in arXiv:1111.4076 criticizing Ref. \onlinecite{sca} (arXiv:1110.0816)
before it was published. These points were clarified in Ref. \cite{com},
but a criticism of Mu\~{n}oz {\it et al.} regarding Ward identities (addressed below)
persists \cite{reply}.
 
On the other hand, my main criticism \cite{com} is that lesser and greater quantities
in Ref. \onlinecite{mu} are incorrect
even in the PHS case. This includes the expressions
for the lesser self energy and Green function, relating them with
the retarded ones
$\Sigma _{{\rm MBK}}^{<}(\omega )=2if_{{\rm eff}}(\omega )$Im$[\Sigma ^{r}(\omega )]$, 
$G _{{\rm MBK}}^{<}(\omega )=-2if_{{\rm eff}}(\omega )$Im$[G^{r}(\omega )]$, 
where $f_{{\rm eff}}(\omega )$ is the average of the Fermi
function at the two leads, weighted by the corresponding $\Gamma _{\nu }$.
The hybridization term, leading to the broadening $i \Delta$ can be included
either in the non interacting  Hamiltonian or in the perturbation. 
Clearly, the first (simpler) approach
was followed \cite{mu}. The retarded quantities are right in the PHS case, 
but the lesser quantities are not, as can be checked comparing with earlier work
on PTU \cite{her}. 

As explained in Ref. \onlinecite{com}, out of the
PHS case, if one uses incorrect lesser and greater quantities, conservation of the current
is not guaranteed, and even retarded quantities might be wrong. The rapid 
deterioration of the agreement with numerical-renormalization-group (NRG) 
results at equilibrium as the system is moved away from PHS is suggestive \cite{merker}.

The arguments in Ref. \onlinecite{reply} regarding an alleged failure 
of Ward identities in the analytical expression for $\Sigma^{<}(\omega )$
given by Eq. (20) of Ref. \onlinecite{sca} is flawed because the authors
assumed that $\partial \Sigma^{<} / \partial V$ is an analytic function, 
which is not true at $T=0$. To keep the argument simple, let us consider 
symmetric voltage drops and couplings to the leads (this implies $\gamma=0$
in the notation of Ref. \onlinecite{sca} which I use here). In this case, 
Eq. (20) of Ref. \onlinecite{sca} coincides with Eq. (87) of Ref. \onlinecite{her} at $T=0$.
Trivial derivation leads to 

\begin{eqnarray}
\frac{\partial \tilde{\Sigma}^{<}}{\partial eV} &=&
-\frac{3}{8}i\pi [\widetilde{\rho }_{0}(0)]^{3} \widetilde{U}^{2} (a_3+a_1 -a_{-1}-a_{-3}),\nonumber \\
a_j(V,\omega)&=&\theta (jeV/2-\omega)(jeV/2-\omega).  
 \label{dsl}
\nonumber \\
\end{eqnarray}
Obviously, for $V$=0, $\partial \Sigma^{<} / \partial V$ vanishes identically, as required
by the Ward identity 
$\partial \Sigma^{<} / \partial eV = -\gamma (\partial \Sigma^{<} / \partial \omega + 
\partial \Sigma^{<} / \partial E_d$) \cite{ogu}. This identity is valid only for $V=0$ 
since the arguments used to relate the different derivatives require that
the Fermi levels of both leads coincide. Due to the non analyticity of the step functions $\theta (\omega )$,
even a tiny $V$ leads to a term proportional to $\omega$ for $\omega \rightarrow 0$ 
in Eq. (\ref{dsl}). 
This term is quite right. 
This points that it is at least dangerous to use Ward identities for an expansion 
of the self energies around $T=0$ as done in Ref. \onlinecite{mu}.

For the general asymmetric case, derivation of Eq. (20)  of Ref. \onlinecite{sca}
and evaluation at $V=0$ leads to (up to terms linear in $\omega$) \cite{note}

\begin{eqnarray} 
&&\frac{\partial \tilde{\Sigma}^{<}(\omega )}{\partial \omega} =
-2i\pi [\widetilde{\rho }_{0}(0)]^{3} \widetilde{U}^{2} \omega \theta(- \omega)  \nonumber \\
&&\frac{\partial \tilde{\Sigma}^{<}(\omega )}{\partial eV} = -\gamma
\frac{\partial \tilde{\Sigma}^{<}(\omega )}{\partial \omega}
\label{dg}
\end{eqnarray}
Since $\partial \Sigma^{<} / \partial E_d$ is of higher order, this also agrees with 
the Ward identity. More details and extension to finite temperature are in Ref. \cite{ng}.  


If $\widetilde{\rho }_{0}(0)$ and  $\widetilde{U}$ are determined from either susceptibility 
and specific heat, NRG or Bethe ansatz, there is no problem with overcounting. 
Terms of higher order in $\widetilde{U}$ lead
to terms of higher order in $\omega$ and $V$.



\begin{thebibliography}{9}
\bibitem{sca} A. A. Aligia, J. Phys. Condens. Matter \textbf{24}, 015306
(2012).

\bibitem{mu} E. Mu\~{n}oz, C. J. Bolech, and S. Kirchner, Phys. Rev. Lett. 
\textbf{110}, 016601 (2013).

\bibitem{com} A. A. Aligia, Phys.  Rev. Lett. \textbf{111}, 089701 (2013).

\bibitem{reply} E. Mu\~{n}oz, C. J. Bolech, and S. Kirchner, Phys. Rev. Lett.
\textbf{111}, 089702 (2013).

\bibitem{her} S. Hershfield, J.?H. Davies, and J.?W. Wilkins, Phys. Rev. B \textbf{46}, 7046 (1992).

\bibitem{merker} L. Merker, S. Kirchner, E. Mu\~{n}oz, and T. A. Costi, Phys. Rev. B \textbf{87}, 165132 (2013).
See also the comment to this work, arXiv:1312.7266, accepted in Phys. Rev. B.

\bibitem{ogu} A. Oguri, Phys. Rev. B \textbf{64}, 153305 (2001).

\bibitem{note} In this case, a printing error of Eq. (20)  of Ref. \onlinecite{sca}
should be corrected. The prefactor of the last term is $\beta_L \beta_R^2$ instead of 
$\beta_L^2 \beta_R$.

\bibitem{ng} A. A. Aligia, Phys. Rev. B 89, 125405 (2014).


\end{thebibliography}
\end{document}